\documentclass{PoS}
\usepackage{layout,amsmath}

\title{Direct photon observables from hydrodynamics and implications on the initial temperature and EoS}

\ShortTitle{Direct photon observables from hydrodynamics}

\author{\speaker{M\'at\'e~Csan\'ad}\\
        E\"otv\"os University,  Department of Atomic Physics, P\'azm\'any P\'eter s. 1/A, H-1117 Budapest, Hungary\\
        E-mail: \email{csanad@elte.hu}}

\abstract{
The expansion of the strongly interacting quark gluon plasma (sQGP) created in Au+Au collisions at RHIC can be described by
hydrodynamical models. Hadrons are created after a freeze-out, thus their distribution describes the final state
of the evolution. The earlier stages can be analyzed via penetrating probes like photon observables. These were
measured in 2010 and 2011 by the PHENIX experiment. Here we analyze an analytic, 1+3 dimensional perfect relativistic 
hydrodynamic solution and calculate hadron and photon observables, such as transverse momentum spectra, elliptic
flow and correlation (HBT) radii. We find that our model is not incompatible with the data, not even with the direct photon
elliptic flow. From fitting the data, we find that early temperatures of the sQGP were well above
the quark-hadron transition temperature, in the hottest point, the center of the fireball the temperature may have
reached 507$\pm$12 MeV. The equation of state of this quark matter can be described by an average sound speed of 0.36$\pm$0.02.
We also predict a photon source that is significantly larger in the \emph{out} direction than in the \emph{side} direction.
}

\FullConference{The Seventh Workshop on Particle Correlations and Femtoscopy\\
		 September 20 - 24 2011\\
		 University of Tokyo, Japan}

\begin{document}
\section{A hydrodynamic model}
It is a well established picture~\cite{Adcox:2004mh} that a strongly interacting quark gluon plasma is created in relativistic
Au+Au collisions of the Relativistic Heavy Ion Collider, and the evolution of this medium can be described by perfect hydrodynamics.
The equations of hydrodynamics can be solved numerically, which has the advantage of having arbitrary initial conditions.
It is also possible to find analytic solutions with realistic properties. However, very few truly 1+3 dimensional (and not spherically
symmetric), relativistic models were compared to data yet. In this paper we analyze the relativistic, ellipsoidally symmetric model
of Ref.~\cite{Csorgo:2003ry}. Hadronic observables were calculated in Ref.~\cite{Csanad:2009wc}, while photonic observables in
Ref.~\cite{Csanad:2011jq}.

The picture used in hydro models is that the pre freeze-out (FO) medium is described by hydrodynamics, and the post freeze-out medium
is that of observed hadrons. The hadronic observables can be extracted from the solution via the phase-space distribution at the
FO. This will correspond to the hadronic final state or source distribution $S(x,p)$. See details about this topic in Ref.~\cite{Csanad:2009wc}.
It is important to see that the same final state can be achieved with different equations of state or initial conditions~\cite{Csanad:2009sk}.
However, as discussed below, the source function of photons is sensitive to the whole time evolution, thus both to initial conditions
and equation of state as well.

For the direct photon calculations~\cite{Csanad:2011jq}, the key assumption is, that even though direct photons may not be thermalized in
the strongly interacting plasma (as their mean free path may be on the order of the size of the fireball), but the radiation itself is thermal.
Thus the phase-space distribution of the photons is characterized by the temperature of the medium (at a given space-time cell), while
the expansion of the fireball also effects the observed spectrum. This is a macroscopic model, and in the following we will calculate photon
observables from it and compare to RHIC data. The most important assumption is, that the spectrum of direct photons is thermal because
macroscopically, the photon radiation is thermal.

The analyzed solution~\cite{Csorgo:2003ry} assumes self-similarity and ellipsoidal symmetry, as described also in
Refs.~\cite{Csanad:2009wc,Csanad:2011jq}. 
The ellipsoidal symmetry means that at a given proper time the thermodynamical quantities are
constant on the surface of expanding ellipsoids.
The ellipsoids are given by constant values of the scale variable $s$, defined as
\begin{align}
s=\frac{r_x^2}{\dot X^2 t^2}+\frac{r_y^2}{\dot Y^2 t^2}+\frac{r_z^2}{\dot Z^2 t^2},
\end{align}
where the constants $\dot X$, $\dot Y$, and $\dot Z$ describe the expansion rate of the fireball in the three
spatial directions. Spatial coordinates are $r_x$, $r_y$, and $r_z$.
The velocity-field is described by an izotropic Hubble-type expansion:
\begin{align}
u^\mu (x) = \frac{x^\mu}{\tau}
\end{align}
where $x$ means the four-vector $(t,r_x,r_y,r_z)$ and $\tau = \sqrt{x^2}$ is the proper-time coordinate.

The temperature distribution $T(x)$ is given as
\begin{align}
T(x)=T_0\left(\frac{\tau_0}{\tau}\right)^{3/\kappa} \exp\left(\frac{b s}{2}\right),\label{e:temp}
\end{align}
where $\tau$ is the proper time, $s$ is the above scaling variable, while $T_0=T|_{s=0,\tau=\tau_0}$,
and $\tau_0$ is an arbitrary proper time, but practically we choose it to be the time of the freeze-out, thus $T_0$ is the central freeze-out temperature.
Parameter $b$ is proportional to the temperature gradient, i.e. if the fireball is the hottest in the center, then $b<0$.
If there is a conserved charge in the system e.g.\ the baryon number density, then charge number density $n(x)$ can
be utilized in the solution. As described in Refs.~\cite{Csorgo:2003ry,Csanad:2009wc}, such a number density can
be introduced as
\begin{align}
n(x)=n_0\left(\frac{\tau_0}{\tau}\right)^3 \exp\left(-\frac{b s}{2}\right).
\end{align}
For the momentum distribution of direct photons, this will not be needed, as the only the temperature of the medium
(the strongly interacting plasma) governs the creation of photons, not the density (which however plays an important
role also in the case of hadron creation). The equation of state (EoS) we use here is $\epsilon = \kappa p$, with
$\epsilon$ being the energy density and $p$ the pressure. Here $\kappa=c_s^{-2}$ (one over speed of sound
squared) is the main parameter describing the EoS.

From the above hydrodynamic quantities, source functions can be created. For bosonic hadrons, it takes the following form~\cite{Csanad:2009wc}:
\begin{align}
S(x,p)d^4x=\mathcal{N}\frac{p_{\mu}\,d^3\Sigma^{\mu}(x)H(\tau)d\tau}{n(x)\exp\left(p_{\mu}u^{\mu}(x)/T(x)\right)-1},
\end{align}
where $\mathcal{N}=g/(2\pi)^3$ (with $g$ being the degeneracy factor), $H(\tau)$ is the proper-time probability distribution
of the FO. It is assumed to be a $\delta$ function or a narrow Gaussian centered at the freeze-out proper-time $\tau_0$. Furthermore,
$\mu(x)/T(x)=\ln n(x)$ is the fugacity factor and $d^3 \Sigma_\mu(x)p^\mu$ is the Cooper-Frye
factor~\cite{Cooper:1974mv} describing the flux of the particles, and $d^3 \Sigma_\mu(x)$ is the vector-measure of
the FO hyper-surface. Here the source distribution is normalized such as $\int S(x,p) d^4 x d^3{\bf p}/E = N$,
i.e. one gets the total number of particles $N$ (using $c$=1, $\hbar$=1 units).

For the source function of photon creation we have~\cite{Csanad:2011jq}:
\begin{align}\label{e:source}
S(x,p)d^4x = \mathcal{N'}\frac{p_{\mu}\,d^3\Sigma^{\mu}(x)dt}{\exp\left(p_{\mu}u^{\mu}(x)/T(x)\right)-1}
= \mathcal{N'}\frac{p_{\mu}u^{\mu}}{\exp\left(p_{\mu}u^{\mu}(x)/T(x)\right)-1}\,d^4x
 \end{align}
where $p_{\mu}d^3\Sigma^{\mu}$ is again the Cooper-Frye factor of the emmission hypersurfaces. Similarly to Ref.~\cite{Csanad:2009wc} we assume that the hyper-surfaces
are parallel to $u^\mu$, thus $d^3\Sigma^{\mu}(x) = u^{\mu}d^3x$. This yields then $p_{\mu}u^{\mu}$ which is the energy of the photon
in the co-moving system. The photon creation is the assumed to happen from an initial time $t_i$ until a point sufficiently near
the freeze-out.

Experimental observables can then be calculated from the source function, using a second order saddlepoint approximation.
In this approximation the point of maximal emissivity is
\begin{align}\label{e:r0}
r_{0,i} = \rho_i t \frac{p_i}{E}\textnormal{ , for }i=x,y,z
\end{align}
while the widths of the particle emitting source are
\begin{align}\label{e:R}
R_i^2 = \rho_i  \tau_0^2 \frac{T_0}{E}\left( \frac{t}{\tau_0}\right)^{-3/\kappa+2}\textnormal{ , for }i=x,y,z
\end{align}
where we introduced the auxiliary quantities
\begin{align}\label{e:rho} 
\rho_i = \frac{\kappa}{\kappa -3-\kappa b/\dot{R_i^2}}
\end{align}
where again $\kappa=c_s^{-2}$ is describing the EoS, and $\dot R_i = \dot X, \dot Y, \dot Z$ for $i=x,y,z$, respectively.

\section{Calculated observables}
The invariant one-particle momentum distribution is defined as $N_1(p)=\int{S(x,p)d^4x}$.
It depends on the three-momentum $p=(p_x,p_y,p_z)$. We will introduce the $(p_t, \varphi, p_z)$ cylindrical coordinates
($z$ being the beam direction) and use the longitudinal rapidity $y$ (for which $E\, dy = dp_z$ is true). As usual, we will
restrict our calculations to $y=0$ (note that in this case $E=p_t$ is true for photons). Our calculated quantities will then be
the elliptic flow $v_2$, and the transverse momentum distribution $N_1(p_t)$. These can be calculated from $N_1(p)$ as
\begin{align}
N_1(p_t) &= \int_0^{2\pi} \! \left.N_1(p)\right|_{y=0} \, d\varphi\\
v_2(p_t) &= \frac{\frac{1}{2\pi}\int_0^{2\pi} \!\left.N_1(p)\right|_{y=0} \cos (2\varphi) \, d\varphi}{N_1(p_t)}
\end{align}

We also calculated Bose-Einstein correlation radii from our model. As usual, the two-particle correlation function
for identical particles can be calculated from the single particle source function $S(x,p)$ as
\begin{align}
C_2(q)=1+\lambda\left|\frac{\widetilde S(q)}{\widetilde S(0)}\right|^2.
\end{align}
where $q$ is the momentum difference of the two particles and $\widetilde S(q)$ is the Fourier-transformed of the source $S(x,p)$
in the variable $x$, and the dependence on the momentum $p$ is not noted in the formulas. This correlation function has, as usual,
a shape with a peak, the width of which is characterized by the HBT radii $R_\textnormal{out}$, $R_\textnormal{side}$ and 
$R_\textnormal{out}$. We calculated these radii for different average momenta $p$.

Here we do not detail the analytic result of these calculation, but will show a comparision of the model to the data in the
next section. The detailed results are given in Refs.~\cite{Csanad:2009wc,Csanad:2011jq}.
It is important to note however, that in the final formulas, we use transverse expansion ($u_t$) and eccentricit
 ($\epsilon$) instead of $x$ and $y$ direction expansion rates $\dot X$ and $\dot Y$:
\begin{align}
\frac{1}{u_t^2} = \frac{1}{2}\left( \frac{1}{\dot{X^2}} + \frac{1}{\dot{Y^2}} \right),
\epsilon = \frac{\dot{X^2}-\dot{Y^2}}{\dot{X^2}+\dot{Y^2}}.
\end{align}

\section{Comparison to the measured direct photon spectrum}
The freeze-out parameters were determined from hadronic fits in Ref.~\cite{Csanad:2009wc}. These properties include the expansion rates,
the freeze-out proper-time and freeze-out temperature (in the center of the fireball), as shown in Table~\ref{t:param}.
When describing direct photon data, we used the parameters of the hadronic fit and left only the remaining as free
parameters~\cite{Csanad:2011jq}. The free parameters will be $\kappa$ (the equation of state parameter) and $t_i$, the initial time of the evolution.

\begin{figure}
	\centering
		\includegraphics[height=0.48\textwidth,angle=270]{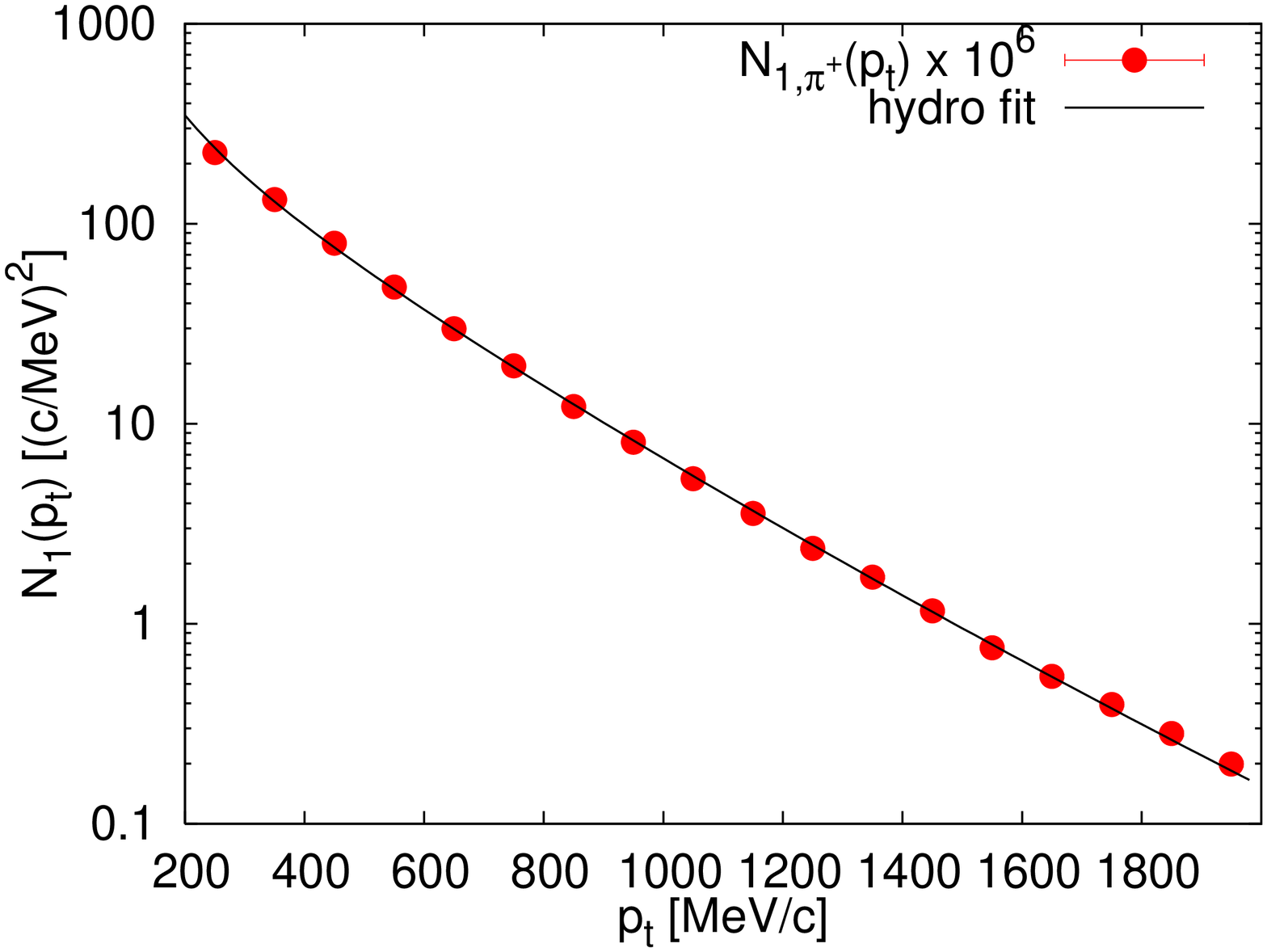}
		\includegraphics[height=0.48\textwidth,angle=270]{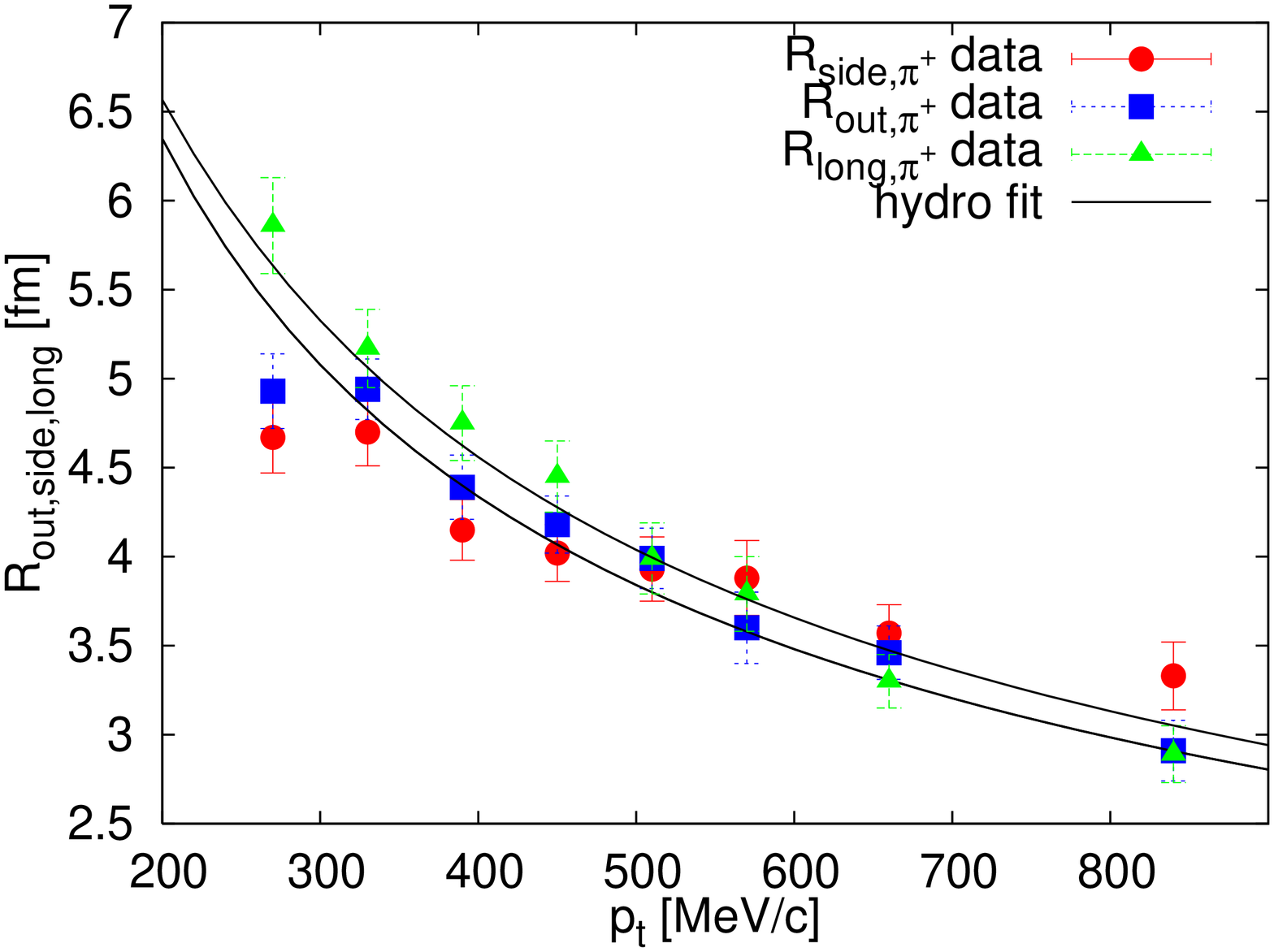}
		\includegraphics[height=0.48\textwidth,angle=270]{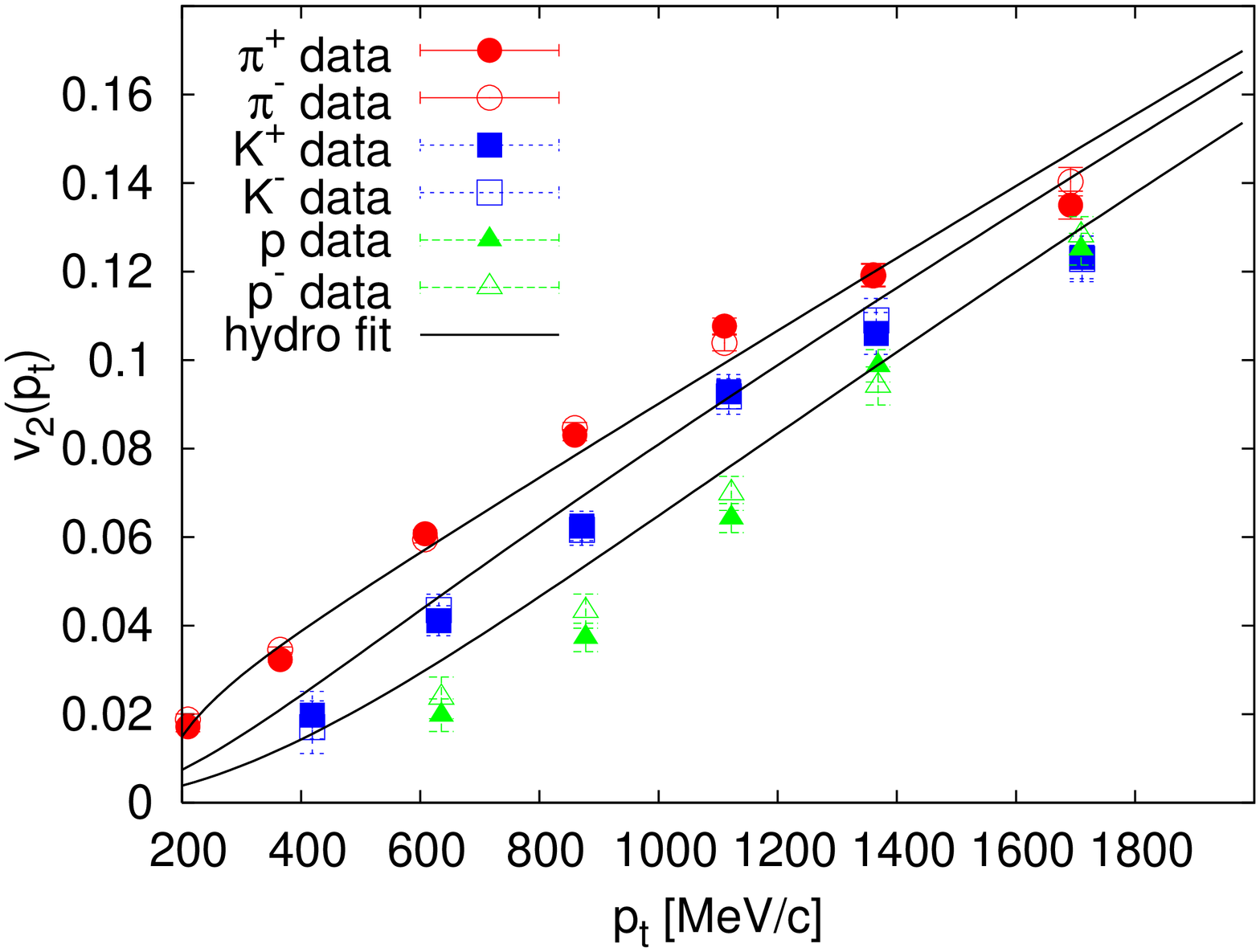}
	\caption{Fits to invariant momentum distribution of pions~\cite{Adler:2003cb} (top left), HBT radii~\cite{Adler:2003kt} (top right)
                         and elliptic flow~\cite{Adler:2004rq} (bottom). See the obtained parameters in
                          Table~\protect\ref{t:param}. In the middle plot the lower curve is the fit to $R_\protect{out}$ and $R_\protect{side}$.}
	\label{f:hadronfits}
\end{figure}

\begin{figure}
 \begin{center}
 \includegraphics[width=0.46\textwidth]{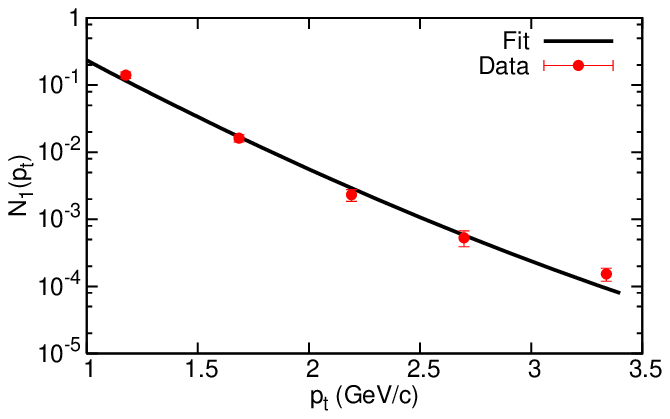}
 \includegraphics[width=0.46\textwidth]{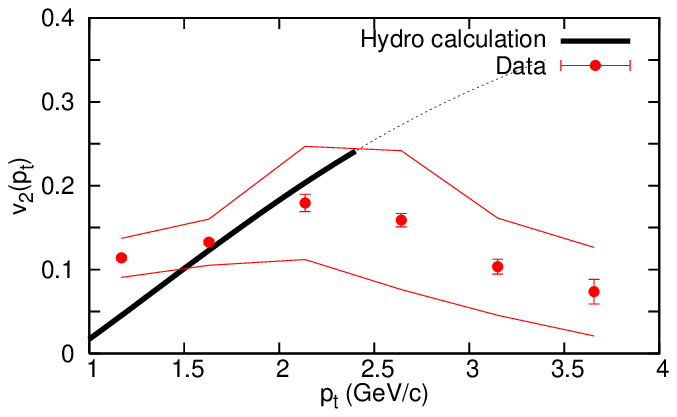}
 \includegraphics[width=0.40\textwidth]{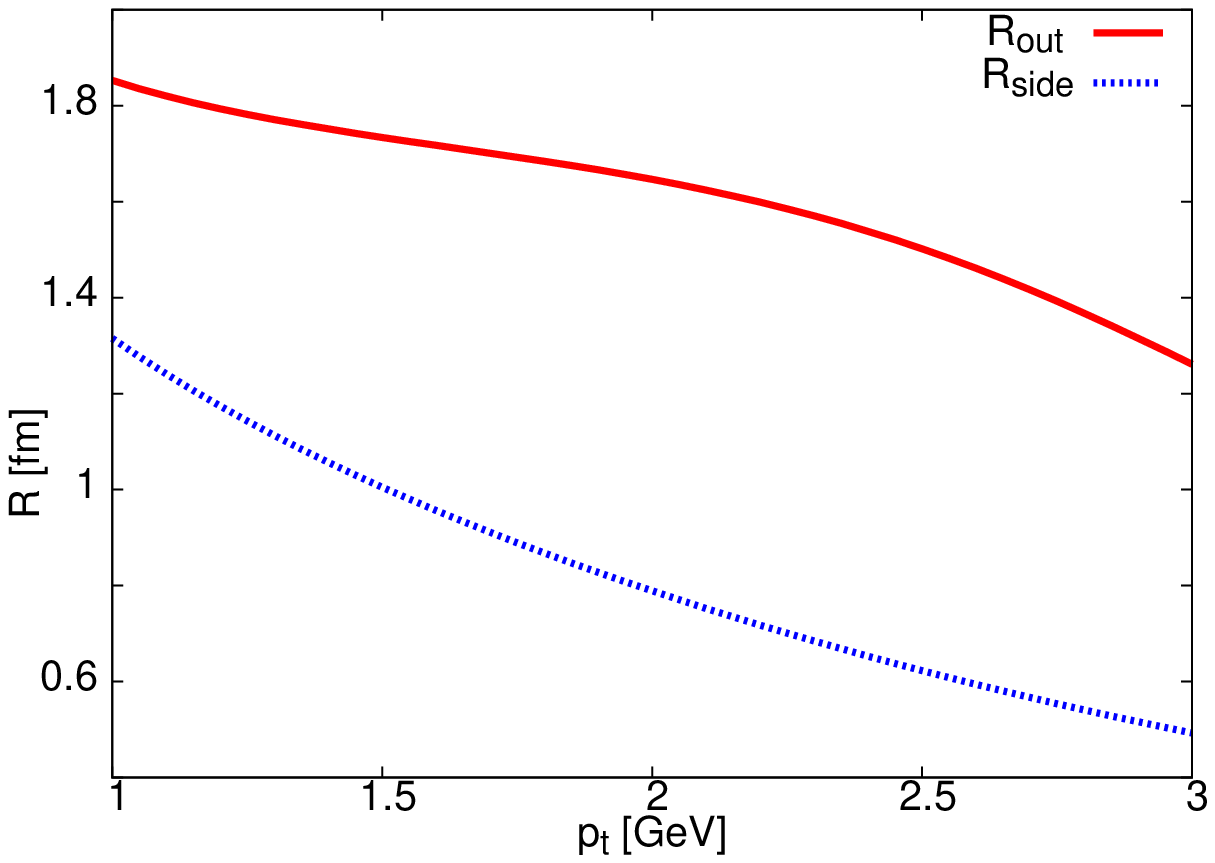}
 \end{center}
 \caption{Fit to direct photon invariant transverse momentum data~\cite{Adare:2008fqa} (top left), comparision
                to elliptic flow data~\cite{Adare:2011zr} (top right) and direct photon HBT predictions (bottom). See the
                model parameters in Table~\protect\ref{t:param}.}\label{f:photonfits}
\end{figure}

We compared the above results to PHENIX hadron and photon data of 200 GeV Au+Au collisions. We fitted our above formulas to
PHENIX invariant transverse momentum distributions of Ref.~\cite{Adler:2003cb}, HBT radii of Ref.~\cite{Adler:2004rq} 
and elliptic flow data of Ref.~\cite{Adler:2003kt}. We used direct photon data also from PHENIX~\cite{Adare:2008fqa,Adare:2011zr}.
Results are shown in Figs.~\ref{f:hadronfits} and~\ref{f:photonfits}, while the model parameters are detailed in Table.~\ref{t:param}.

\begin{table}
\begin{center}
    \begin{tabular}{ lllll }
    \hline\hline
    Dataset                	&                               	& $N_1$ and HBT	& elliptic flow    	& $N_1$\\
                                 	&                              	& 0-30\%  cent.  	& 0-30\% cent.	&  0-92\% cent.\\
                                  	&                              	& hadrons            	& hadrons        	& photons\\\hline
    Central FO temperature	& $T_0$ [MeV]         	& 199$\pm$3      	& 204$\pm$7   	& $204$ MeV (fixed) \\
    Eccentricity                  	& $\epsilon$             	& 0.80$\pm$0.02 	& 0.34$\pm$0.03	& $0.34$ (fixed) \\
    Transverse expansion  	& $u_t^2/b$            	& -0.84$\pm$0.08	& -0.34$\pm$0.01	& $-0.34$ (fixed) \\
    FO proper-time            	& $\tau_0$ [fm/$c$]	& 7.7$\pm$0.1    	& -                   	& $7.7$  (fixed) \\
    Longitudinal expansion 	& $\dot{Z_0^2}/b$   	& -1.6$\pm$0.3  	& -                   	& $-1.6$ (fixed) \\
    Equation of State       	& $\kappa$               	&  -                      	& -                   	& $7.9 \pm 0.7$  \\
    Initial time                     	& $t_i$  [fm/$c$]       	&  -                      	& -                   	& $0 - 0.7$ fm$/c$ \\ \hline 
    Fit quality                   	&                               	&                       	&                     	&  \\ \hline
    Degrees of freedom   	& NDF                           	& 41                    	& 34                     	& 3\\ 
    Chisquare                  	& $\chi^2$               	& 24                   	&  66                      	& 7\\ 
    Confidence level        	&                              	& 98\%                 	&  0.1\%            	& 7.2\%\\
    \hline\hline
    \end{tabular}   
\end{center}
\caption{Parameters of the solution, describing the expanding sQGP. The first five were determined from hadronic fits~\cite{Csanad:2009wc},
              the remaining from direct photon data~\cite{Csanad:2011jq}. See details in the text and in these references.}\label{t:param}
\end{table}

Let us discuss the results of the photon fits! The equation of state result is $\kappa=7.9\pm0.7$, or alternatively,
using $\kappa=1/c_s^2$:
\begin{align}
c_s =  0.36\pm0.02_{stat}\pm0.04_{syst}
\end{align}
which is in nice agreement with both lattice QCD calculations~\cite{Borsanyi:2010cj} and experimental
results from hadronic data~\cite{Adare:2006ti,Lacey:2006pn}. This represents an average EoS as it may
vary with temperature. There may be solutions with a $\kappa(T)$ function~\cite{Csanad:2012xx}, but for the sake of simplicity we
assumed here an average, fixed $\kappa$. As detailed in Ref.~\cite{Csanad:2011jq}, we
determined an ``interval of acceptability'' for $t_i$. The maximum value for $t_i$ within 95\% probability is 0.7 fm/$c$. This can then be used
to determine a lower bound for the initial temperature, using the eq.~\ref{e:temp}. Thus the initial temperature of the fireball (in its center) is:
\begin{align}
T_i > 507\pm12_{stat}\pm90_{syst}\textnormal{MeV}
\end{align}
at 0.7 fm/$c$. This is in accordance with other hydro models as those values are in the $300-600$ MeV interval~\cite{Adare:2008fqa}.
Note that a systematic uncertainty was determined by using a prefactor of $(T/T_0)^N$, with $N=0,1,2,3$.
This factor arises if the photon creation can be described by a microscopic process, as detailed in Ref.~\cite{Csanad:2011jq}.
This causes only a minor change in the resulting spectrum, as it is dominated by the exponential factors in it.
However, the equation of state parameter $\kappa$ changes from 7.9 to 6.5 as we increase the exponent in the prefactor.
This gives a systematic uncertainty to our parameters.

A measurement of direct photon elliptic flow was also performed recently at PHENIX~\cite{Adare:2011zr}. Using the previously
determined fit parameters we can calculate the elliptic flow of direct photons in Au+Au collisions at RHIC.
Due to the low number of points in the desired range, a fit could not be performed here, but we used the
average value $\epsilon$ in case of the two fits of Ref.~\cite{Csanad:2009wc}.
The resulting curve, where the value $\epsilon=0.59$ was used, is shown on Fig.~\ref{f:photonfits}.

In case of hadronic HBT, correlation radii in the side and out directions are almost equal, as for the hadronic transition is of
cross-over type (i.e. the transition time is short), see details in Ref.~\cite{Csanad:2009wc}. However, in case of
photons, the creation spans the whole evolution of the fireball, thus $R_\textnormal{out}$ will be significantly larger than
$R_\textnormal{side}$. Indeed this was observed in our model, as shown on Fig.~\ref{f:photonfits}. 

To summarize, we find that thermal radiation is consistent with direct photon data, and our result on the equation of state
is $c_s = 0.36\pm0.02_{stat}\pm0.04_{syst}$. We set a lower bound on the initial temperature of the sQGP to $507\pm12_{stat}\pm90_{syst}$ MeV at $0.7$ fm/$c$. We also find that the thermal photon
elliptic flow from this mode is not incompatible with measurements. We also predicted photon HBT radii from the model, and
discovered a significantly larger $R_\textnormal{out}$ than $R_\textnormal{side}$.

\section*{Acknowledgments}
The author would like to thank the invitation to the WPCF 2011 conference, and the kind hospitality of the organizers, in particular T. Hirano. The author also would like to thank T. Cs\"org\H{o} and S. Pratt for valuable discussions and important insigths on the topic of
this paper. M. Csan\'ad gratefully acknowledges the support of the Bolyai Scholarship of the Hungarian Academy of Sciences.

\bibliographystyle{h-physrev}
\bibliography{../../../master}

%

\end{document}